# Magnetic charge's relaxation propelled electricity in two-dimensional magnetic honeycomb lattice

Yiyao Chen[1], George Yumnam[1], Jiasen Guo[1], Laura Stingaciu[2], Piotr Zolnierczuk[2,3], Valeria Lauter[2], and Deepak K. Singh[1,4]*

[1]Department of Physics and Astronomy, University of Missouri, Columbia, MO 65211, USA
[2]Neutron Scattering Division, Oak Ridge National Laboratory, Oak Ridge, TN 37831, USA
[3]Forschungszentrum Julich GmbH, JCNS Outstation at SNS, Oak Ridge National Laboratory, Oak Ridge, TN 37831, USA
[4]Lead Contact
*Correspondence: singhdk@missouri.edu

**SUMMARY**

Emerging new concepts, such as magnetic charge dynamics in two-dimensional magnetic material, can provide novel mechanism for spin based electrical transport at macroscopic length. In artificial spin ice of single domain elements, magnetic charge's relaxation can create an efficient electrical pathway for conduction by generating fluctuations in local magnetic field that couple with conduction electrons spins. In a first demonstration, we show that the electrical conductivity is propelled by more than an order of magnitude at room temperature due to magnetic charge defects sub-picosecond relaxation in artificial magnetic honeycomb lattice. The direct evidence to the proposed electrical conduction mechanism in two-dimensional frustrated magnet points to the untapped potential for spintronic applications in this system.

**INTRODUCTION**

Electrical transport phenomenon is a core physical property of materials. Unlike the well understood metal or semiconductor (Kasap et al., 2017), there are numerous other materials, ranging from magnetic insulator to biological systems, where the underlying mechanism behind electricity propagation is still not understood (Brataas et al., 2012, Hu 2012). For instance, magnon (spin wave) propagation is argued to be the electricity transportation medium in magnetic insulator (Cornelissen et al., 2015). Similarly, proton-based conduction mechanism is proposed to explain electric current flow in biological materials (Ordinario et al., 2014). The unconventional transport mechanism due to the coupling between magnetic moment dynamics and charge's spin is touted to be an efficient electrical transmission process, albeit at low temperature, in solid state devices (Kajiwara et al., 2010; Zutic et al., 2004). Spin based conduction, as envisaged in spintronics, is an emergent research arena with immense practical implications (Chappert et al., 2007; Zutic et al., 2004). Recently, an entirely different electrical effect due to magnetic monopoles dynamics was proposed in spin ice magnet with peculiar two-in & two-out spin structure on a tetrahedron motif (Bramwell et al., 2009; Bramwell 2012). According to the Dumbbell model's prescription (Castelnovo et al., 2008; Gardner et al., 2010), a magnetic moment is made of magnetic Coulomb's interaction bound two equal and opposite magnetic charges, ±Q (A.m). In spin ice, the flipping of a moment's direction due to external effect creates a charge defect (monopoles) $Q_m$ of ±2Q unit, which traverses the lattice with very little energy loss (Bramwell 2012). Arguably, monopole's dynamics generates non-Maxwellian electricity in applied magnetic field at low temperature, T <15 K, such that $\nabla \cdot \boldsymbol{B} \neq 0$ (Bramwell et al., 2009; Castelnovo et al., 2008; Dunsiger et al., 2011). However, monopole electricity, directly arising due to the monopole's motion, is too weak to be measured directly and is typically inferred from the μSR measurement (Bramwell et al., 2009). Monopole's current need not be confused with conventional electric current. A recent theoretical research has suggested that monopole or magnetic charge defect dynamics causes transverse fluctuation in local magnetic field, which can interact with electric charge carrier (Chern et al., 2013). The indirect interaction between monopoles dynamics and electric charge carrier can lead to the efficient electrical conduction process. A necessary condition to realize the magnetic charge dynamics mediated conduction requires that the charge defect dynamics be confined along a particular direction. Artificial spin ice, such as artificial magnetic honeycomb lattice, with narrow connecting element automatically fulfills this criterion. An outstanding question is: can dynamic charge defect spur an electrical conduction process at high temperature? We have answered this question in the article.

An artificial spin ice, such as a nanoengineered honeycomb lattice made of single domain magnetic (permalloy, $Ni_{0.81}Fe_{0.19}$) elements (typical length l ~ 10 nm), provides a facile platform to explore electrical conduction mediated by magnetic charge defect at higher temperature. The flexibility in tuning constituting element's size, hence the inter-elemental dipolar interaction energy, in an artificial magnetic honeycomb lattice can be exploited to generate magnetic charge defects at high temperature. Previously, magnetic charge defect was shown to mediate electrical conduction in a spin ice material via the interaction between electron's spin, **s**, at the Fermi surface and the transverse local fluctuation in magnetic field **B(k)** due to the monopoles dynamics (Chern et al., 2013; Lopez-Bara et al., 2017). According to the formulation, magnetic charge defect's relaxation time plays crucial role in temperature dependent magnetic charge mediated conduction, in addition to the residual purely electrical conduction. Following that, a slightly modified equation for conductivity can be written as:

$$\sigma = \sigma_0 / \left(1 - A.h \frac{e^{-\Delta/k_B T}}{\varepsilon_F \tau_m}\right) \quad (1)$$

where we use $\sigma_0$ as the residual electrical conductivity, given by the Drude's formula $\frac{ne^2\tau_e}{m}$. The dimensionless constant $A$ ($\sim \kappa. \frac{k_F}{l_c^{-1}}$) depends on the Fermi wave vector of permalloy, the cut-off length $l_c$, limited by the thickness of the lattice, for transverse magnetic field fluctuation and a temperature independent constant factor, $\kappa$, due to magnetic lattice. Magnetic charge mediated conductivity also depends on two crucial analytical parameters: $\Delta$ and $\tau_m$ that represent the threshold energy to create magnetic charge defect and magnetic charge defect's relaxation time, respectively. A key criterion to the applicability of the above formulation (originally envisaged for bulk spin ice materials) requires that the relaxation time, $\tau_m$, in artificial honeycomb lattice be comparable to the spin relaxation time in spin ice magnet. In a honeycomb lattice with two-types of magnetic charges, ±3Q and ±Q associated to the all-in or all-out and 2-in & 1-out (or vice-versa) magnetic moment correlations (Nisoli et al., 2013; Skjaerrvo et al., 2020), respectively, the threshold energy Δ to create magnetic charge defect ±2Q is given by the energy difference |$E_{3Q}$ – $E_Q$| or, |$E_Q$ – $E_{-Q}$|. Here, Q is directly related to magnetic moment M via Q = M/l (Nisoli et al., 2013). Additionally, the unidirectional motion of magnetic charge defect, confined along the length of single domain connecting element due to narrow width and small thickness, can only create transverse fluctuation in local magnetic field i.e. perpendicular to the direction of motion. While the Fermi surface properties of permalloy ($\varepsilon_F$ and $k_F$) in both the bulk and the thin film are well-known (Mijnarends et al., 2002; Petrovyky et al., 1998), the complete lack of information about $\tau_m$ forbids any meaningful deduction of electrical conductivity mediated by magnetic charge defect. We have resolved this issue. We precisely determine $\tau_m$ using neutron spin echo measurements on a parallel stack of permalloy honeycomb lattice samples. The estimated relaxation time of magnetic charges in permalloy honeycomb, ~ 0.05 ns, is comparable to the ~ picosecond spin relaxation in bulk spin ice (Ehlers et al., 2004), which validates the applicability of Equation (1) to two-dimensional system. We use the newly acquired knowledge of $\tau_m$ to demonstrate the significant enhancement in electrical conductivity due to magnetic charge's relaxation in honeycomb lattice.

## RESULTS AND DISCUSSION
### Magnetic charge defect relaxation in artificial honeycomb lattice
In a honeycomb lattice, magnetic charges on the vertices undergo transformation by releasing or absorbing charge defect of 2Q unit magnitude (Nisoli et al., 2013; Skjaerrvo et al., 2020). Previously, researchers have used the electron-beam lithography technique to create two-dimensional artificially frustrated structure (Skjaerrvo et al., 2020). However, the large element size of the 2D lattice makes it 'athermal' as the inter-elemental dipolar interaction energy, typically of the order of $10^4$ K, is much large to achieve in a laboratory (Nisoli et al., 2013; Rougemaille et al., 2011). We overcome this obstacle by utilizing a hierarchical nanofabrication scheme using the diblock template synthesis method, which results in large size honeycomb lattice sample (~ sq. inch) with truly nanoscopic connecting element of permalloy ($Ni_{0.81}Fe_{0.19}$) magnet in the single domain limit ~ 12 nm (length) × 5 nm (width) × 8 nm (thickness) (see Supplemental Information) (Chen et al., 2019; Glavic et al., 2018). Consequently, the inter-elemental dipolar energy in the new lattice is very small, ~ 15 K. An important advantage of small energy scale lies in the feasibility of populating the honeycomb vertices using both ±Q and ±3Q charges at finite temperature. The high integer charges, ±3Q, are usually accompanied by the high energy cost. Hence, they are not stable. To attend a stable or quasi-stable state of Q or -Q charge, the high integer charges emit or absorb the charge defect of magnitude 2Q unit. The charge defect is highly mobile and traverses the lattice. A schematic description of magnetic charge dynamics in the thermally tunable honeycomb lattice is shown in Figure 1A-D. The charge defect can travel between the nearest neighbors or, between the next nearest neighbors, until it faces a high integer charge (3Q or -3Q), which serves as the roadblock. At higher temperature where thermal fluctuation is stronger than $\Delta = |E_Q - E_{-Q}|$ (~ 30 K), the flipping of a magnetic moment associated to the 2-in & 1-out or vice-versa configuration can also release a charge defect of 2Q unit magnitude. The typical relaxation length of magnetic charge defect corresponds to q ~ 0.06 Å$^{-1}$ (between neighboring vertices, Figure 1B) and q ~ 0.03 Å$^{-1}$ (between next nearest neighbor vertices, Figure 1C) in reciprocal space.

We indeed observe strong evidences to magnetic charge defect's relaxation at localized q values of ~ 0.06 Å$^{-1}$ and ~ 0.03 Å$^{-1}$ in neutron spin echo (NSE) measurements, see Figure 1E. NSE is a quasi-elastic measurement technique where the relaxation of magnetic specimen is decoded by measuring relative change in scattered neutron's polarization via the change in the phase current at a given Fourier time (related to neutron precession). The use of neutron spin echo technique to elucidate magnetic dynamic properties in magnetic thin film is not common due to weak signal-to-background ratio. To overcome this, we carried out NSE measurements in a modified instrumental configuration at the Spallation Neutron Source (SNS)-NSE spectrometer at Oak Ridge National Laboratory (ORNL). Here, magnetism in the sample is used as a π flipper to apply 180° neutron spin inversion, instead of utilizing a flipper before the sample (Zolnierczuk et al., 2019). Such a modification not only ensures that the detected signal is magnetic in origin, but also reduces incident neutron intensity loss. The NSE measurements were performed on a parallel stack of 117 samples of 20×20 mm$^2$ size to obtain the good signal-to-background ratio. As shown in Figure 1E, the two-dimensional color plot of scattered magnetic intensity for spin up neutron polarization reveals significant spectral weights at the above mentioned localized q positions. At the same time, little or no spectral weight is detected in the spin down neutron polarization in Figure 1F, confirming the magnetic nature of the signal (also see Figures S1 and S2).

The relaxation of magnetic charge defect is duly affected by the occupation density of energetic ±3Q charges. Before determining the relaxation time of charge defects, we quickly show evidence to the occurrence of high integer charges in our honeycomb lattice using polarized neutron reflectometry (PNR) (Lauter et al., 2016). Figure 2A depicts the atomic force micrograph of a typical honeycomb lattice, created using the diblock templating method (see Supplemental Information for detail). PNR measurements were performed on ~ 1 sq. inch size sample in a small guide field of H = 20 Oe to maintain the polarization of incident and scattered neutrons. In Figure 2B, we plot the off-specular intensity measured using spin-up (+) and spin-down (-) neutron at T = 5 K, obtained on MagRef instrument at SNS. The specular reflectivity lies along the x = 0 line in Figure 2B. While the asymmetry between '+' and '-' components in specular data (see Figure S5) infers magnetism in the honeycomb lattice, a broad band of diffuse scattering along the $q_x$ direction suggests in-plane correlation of magnetic charges on honeycomb vertices (Lauter et al., 2016). Experimental data is modeled using the distorted wave Born approximation (DWBA) formulation (see Supplemental Information) to understand the nature of charge correlation (Glavic et al., 2018). As shown in Figure 2D, the numerically simulated reflectometry pattern for magnetic charge configuration, comprised of both ±Q and ±3Q charges (shown in Figure 2C), is found to be in good agreement with experimental data. The PNR measurements and the associated DWBA modeling basically suggest that a significant number of honeycomb vertices are indeed occupied by ±3Q charges at T = 5 K. At higher temperature, the system is expected to maintain the same or higher density of high integer magnetic charges to generate the sufficient number of relaxing charge defects (Ladak et al., 2012).

**Determination of the relaxation time of magnetic charge defects**

Next, we quantitatively determine the relaxation time of magnetic charge defect, $\tau_m$, in permalloy honeycomb lattice by analyzing the NSE data at various Fourier times and temperatures. Strong sinusoidal oscillations, reminiscent of high quality spin echo, are observed in NSE measurements at T = 4 K, 15 K, 50 K and 300 K. Characteristic spin echo plots at T = 4 K and T = 300 K at q ~ 0.06 Å$^{-1}$ and at a Fourier time of t = 0.1 ns are shown in Figure 3A-B. In Figure 3C-D, we show the plots of normalized intensity, S(q, t)/S(q, 0), as a function of the spin echo Fourier time at q ~ 0.06 Å$^{-1}$ and 0.03 Å$^{-1}$ at T = 300 K. The normalization is achieved by dividing the observed oscillation amplitude by the maximum measurable amplitude (see Supplemental Information), which is a common practice in magnetic systems with unsettling fluctuation to the lowest measurement temperature (Zolnierczuk et al., 2019). The normalized intensity reduces to the background level above the spin echo's Fourier time of ~ 0.5 ns. This is the most general signature of relaxation process in NSE measurements (Ehlers et al., 2006; Zolnierczuk et al., 2019). Fitting of NSE scattering intensity using the typical exponential function yields magnetic charge defect's relaxation time. It is given by (Ehlers et al., 2006),

$$\frac{S(q,t)}{S(q,0)} = C \exp(-\frac{t}{\tau_m}) \tag{2}$$

where C is constant fitting parameter. While the fluctuation is prominent even at T = 4 K in honeycomb sample, the slowing down is considerably faster at T = 300 K, see Figure 3E-F. The exponential function well describes the relaxation mechanism of magnetic charges. The obtained value of $\tau_m$ = 0.049 ns at T = 300 K corresponds to the average magnetic charge defect's velocity of ~ 200 m/s. It is at least three times faster than the estimated domain wall velocity in nanostructured magnetic materials (Zhang and Li 2004). At T = 4 K, $\tau_m$ increases to 0.13 ns. $\tau_m$ at T = 4 K and T = 300 K act as limits for relaxation at intermediate temperatures.

### Electrical conductivity mediated by magnetic charge defects

Finally, we test the hypothesis of electrical conduction mediated by magnetic charge defects in our honeycomb lattice. Although, the probability of magnetic charge defect relaxing along the current application direction is 50%, the electrons will drift opposite to the electric field application and interact with magnetic charges in that particular direction. Two conducting processes take place simultaneously: a, electrons relaxing themselves with finite drift velocity and b, the drifting electron's spin interacts with the fluctuation in magnetic field due to magnetic charge defects dynamics. Since magnetic charge defect's velocity is significantly higher than electron's drift velocity (typically few μm/s), we can expect to observe a weak background contribution to electrical conductivity due to the mechanism (a). We estimate the dc electrical conductivity of permalloy honeycomb from I-V measurements. Plots of I-V traces at characteristic temperatures are shown in Figure 4A. In Figure 4B, we show the plot of estimated conductivity at different temperatures. There is a noticeable non-linearity in the I-V curve as temperature is reduced below T = 300 K. The small slope at low voltage is most likely associated to the background conductivity due to the purely electric charge carrier's relaxation. As temperature reduces, the background conductivity starts becoming prominent. At low temperature, T < 30 K, the background conductivity becomes comparable to the estimated conductivity due to magnetic charge relaxation.

Since $\tau_m$ varies as a function of temperature, unique fit to the conductivity data cannot be obtained. Rather, we generate curves for $\tau_m$ at T = 4 K and T = 300 K using Equation (1). For this purpose, we use the standard values of $k_F$ ~ 1.05 Å$^{-1}$ and $\varepsilon_F$ ~ 0.2 eV for permalloy at low temperature (Mijnarends et al., 2002; Petrovyky et al., 1998). Fermi surface values at higher temperature is corrected using the Fermi-Dirac statistics. The background conductivity due to electron's relaxation, $\sigma_0$ ~ 5 Ω$^{-1}$ cm$^{-1}$, is kept constant at both temperatures. We see that the conductivity, $\sigma$, lies within the two limits imposed by the slower and faster relaxation times at T = 4 K and 300 K, respectively. At room temperature, the conductivity is boosted by more than an order of magnitude with respect to the background conductivity. While the conductivity at higher temperature fits very well to the $\tau_m$ (T = 300 K) curve, at lower temperature the conductivity deviates towards the $\tau_m$ (T = 4 K) curve. At intermediate temperature where $\tau_m$ (300 K) < $\tau_m$ (T) < $\tau_m$ (4 K), $\sigma$ lies between the two limiting curves. It clearly demonstrates the role of new electrical conduction mechanism in 2D honeycomb lattice.

### Conclusions

In summary, we have presented detailed investigation of magnetic charge's relaxation process in two-dimensional artificial permalloy honeycomb lattice. For the first time, neutron spin echo measurement technique is utilized to extract sub-ns relaxation of magnetic charge defect dynamics in an artificial spin ice. More importantly, we have also showed that magnetic charges remain highly dynamic to the lowest measurement temperature in artificial permalloy honeycomb lattice of single domain elements. The ultra-small element size imparts a thermally tunable characteristic to the lattice, which is crucial to the manifestation of magnetic charge mediated electrical conduction process.

Our comprehensive study reveals new electrical conduction process in two-dimensional frustrated magnet. Here, we have shown that the electrical conductivity gets a major boost due to magnetic charge's relaxation in a honeycomb lattice. The much stronger conductivity is attributed to charge defect's unidirectional relaxation (Tomasello et al., 2019), which causes a net drag on the electric charge carriers drift motion via indirect interaction with electron's spin. Spin-based electrical conduction at room temperature has been a major challenge, crucial to the development of spintronic system and devices (Zutic et al., 2004). Artificially frustrated geometry can prove a gamechanger in this quest. Our study also establishes that the modified NSE technique can be utilized to investigate magnetization relaxation properties in a multitude of materials, including magnetic thin film and nanomagnetic devices, that are seemingly difficult at present. Strong understanding of magnetic relaxation in nanodevices (Sun et al., 2005) is highly desirable for technological applications.

## LIMITATIONS OF THE STUDY

To understand the role of magnetic charge mediation in electrical conduction in honeycomb lattice, we have utilized a theoretical formalism that was originally envisaged for the bulk spin ice material. Although the formalism explains the experimental observation of propelled conduction, but a more appropriate theoretical mechanism for artificial spin ice is needed to better understand the phenomenon. There could also be a new possible mechanism behind the propelled electrical conduction in artificial permalloy honeycomb lattice, which we don't know. Our study is limited in that perspective.

## EXPERIMENTAL PROCEDURES

### Resource Availability

#### Lead Contact

Further information and requests for resources and reagents should be directed to and will be fulfilled by the Lead Contact, Deepak K. Singh (singhdk@missouri.edu).

#### Materials Availability

All unique reagents generated in this study are available from the Lead Contact with a completed Materials Transfer Agreement.

#### Data and Code Availability

The published article includes all datasets generated or analyzed during this study.

### Methods

All methods can be found in the accompanying Transparent Methods supplemental file.

## ACKNOWLEDGMENTS

We thank Antonio Faraone and Michihiro Nagao for helpful discussion. D.K.S. thankfully acknowledges the support by the Department of Energy, Office of Science, Office of Basic Energy Sciences under the grant no. DE-SC0014461. This work utilized the facilities supported by the Office of Basic Energy Sciences, US Department of Energy.

## AUTHOR CONTRIBUTIONS

D.K.S. conceived and supervised every aspect of research. Y.C. and G.Y. made equal contributions to the research. Y.C. and G.Y. made the samples. Y.C., G.Y., J.G., L.S., P.Z., V.L. and D.K.S. performed neutron scattering experiments. Y.C. and G.Y. analyzed and performed numerical modeling of NSE and PNR data. Y.C., G.Y. and J.G. performed electrical measurements. D.K.S. prepared the manuscript with input from all co-authors.

**DECLARATION OF INTERESTS**

The authors declare no competing interests.

**Figure 1. Magnetic charge defect's relaxation in an artificial magnetic honeycomb lattice.**
(A) Magnetic charges of ±Q unit and ±3Q unit arise due to local magnetic moment arrangements, aligned along the length of connecting element due to shape anisotropy, on the vertices of the honeycomb lattice. At finite temperature, significant number of vertices in a honeycomb lattice made of single domain size elements (~ 10 nm in length) are occupied by high integer, ±3Q unit, charges. (B-D) High integer charges attend the quasi-stable configuration by releasing or absorbing the magnetic charge defect of 2Q unit magnitude. The charge defect relaxes by traversing the length between nearest neighbors (part B) or between the next nearest neighbors (part C). If two charge defects are released in the same direction or a charge defect is released towards a high integer charge of the same polarity, then relaxation is limited to less than half of the length (part D). (E-F) Experimental evidence to magnetic charge defect's relaxation between nearest and next nearest neighbor vertices via neutron spin echo (NSE) spectroscopy. Each pixel in the color plot corresponds to distinguishable q. Part E shows bright intensities at $q_1 \sim 0.06$ Å$^{-1}$ and $q_2 \sim 0.03$ Å$^{-1}$ (near the y-axis) for neutron polarization along +Z axis. There is also faint scattering at x-pixel ~ 18, corresponding to the situation D. The absence of bright intensity for neutron polarization along -Z axis (part F) suggests magnetic nature of scattering. See Figures S1 and S2 for more details of NSE measurement.

**Figure 2. Evidence of finite density of ±3Q charges in thermally tunable artificial permalloy honeycomb lattice.**
(A) Atomic force micrograph of a typical artificial honeycomb lattice. (B) Off-specular plot, shown as sum of spin-up and spin-down neutrons, of polarized neutron reflectometry (PNR) at T = 5 K. (See Figure S5 for specular data of PNR measurement.) We follow the typical plotting convention for the off-specular reflectometry graph i.e. y-axis represents the out-of-plane scattering vector ($q_z = \frac{2\pi}{\lambda}(sin\ \alpha_i\ +\ sin\ \alpha_f)$) and the difference between the z-components of the incident and the outgoing wave vectors ($p_i - p_f = \frac{2\pi}{\lambda}(sin\ \alpha_i - sin\ \alpha_f)$) is drawn along the x-axis, corresponding to the in-plane correlation. (Glavic et al., 2018; Lauter et al., 2016) (C-D) Numerical modeling of experimental data, as shown in part D, for the magnetic charge configuration, comprised of ±3Q (red and blue balls) and ±Q charges on honeycomb vertices, part C, well describes the experimental reflectometry profile in part B.

**Figure 3. Estimation of magnetic charge defect's relaxation time along honeycomb element.**
(A-B) Strong signal-to-background ratio is observed in representative spin echo oscillations at T = 4 K (part A) and T = 300 K (part B) at $q_1 \sim 0.06$ Å$^{-1}$ (see Figure S3 for other temperatures). (C-D) Plot of S(q, t)/S(q, 0) as a function of neutron Fourier time at $q_1 \sim 0.06$ Å$^{-1}$ and $q_2 \sim 0.03$ Å$^{-1}$ at T= 300 K. Sharp slowing down in magnetic relaxation is observed at both q values. (E-F) Plot of S(q, t)/S(q, 0) as a function of neutron Fourier time at $q_1 \sim 0.06$ Å$^{-1}$ at T = 4 K and T = 300 K. Experimental data is fitted with Equation (2) to extract $\tau_m$ at different temperatures. Relaxation is faster at T = 300 K ($\tau_m$ = 0.049 ns) than at T = 4 K ($\tau_m$ = 0.13 ns), also see Figure S4.

**Figure 4. Magnetic charge defect's mediated electrical conductivity in artificial permalloy honeycomb lattice.**
(A) I-V traces at few characteristic temperatures for permalloy honeycomb (for permalloy thin film see Figure S6). Electrical conductivity is extracted via linear fit to the data. (B) Plot of conductivity $\sigma$ vs. temperature. The conductivity increases as temperature increases. We also show theoretically estimated $\sigma$ due to Equation (1) using $\tau_m$ at T = 300 K (red curve) and at T = 4 K (black curve). The value of $\Delta$ used to generate the curves, ~ 28 K, is very close to the theoretically estimated value in honeycomb lattice. Temperature dependent variation in relaxation time forbids a unique fit to the data. Rather, conductivity data lie between the two theoretically generated limiting curves due to relaxation times at T = 300 K and T = 4 K, respectively.

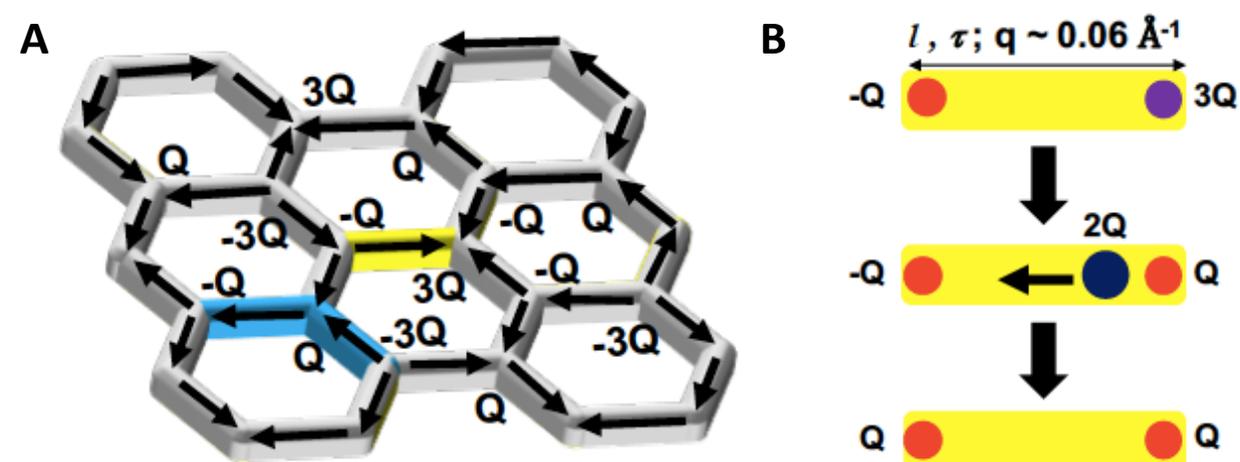
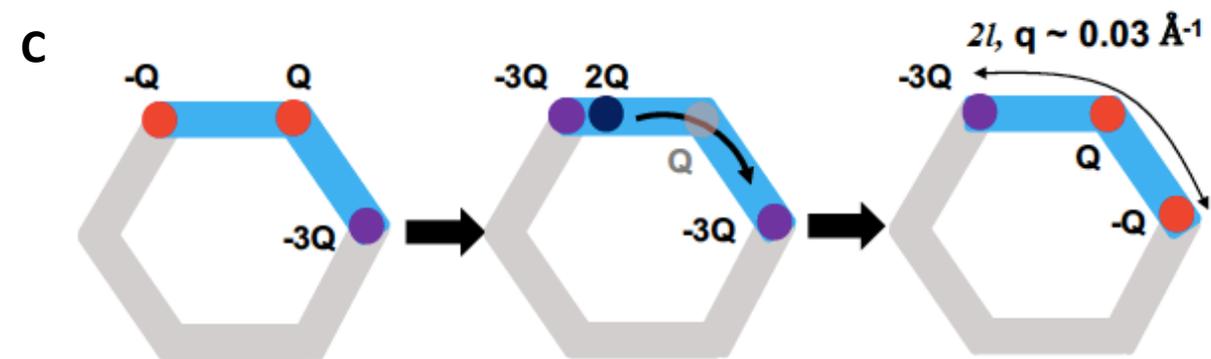
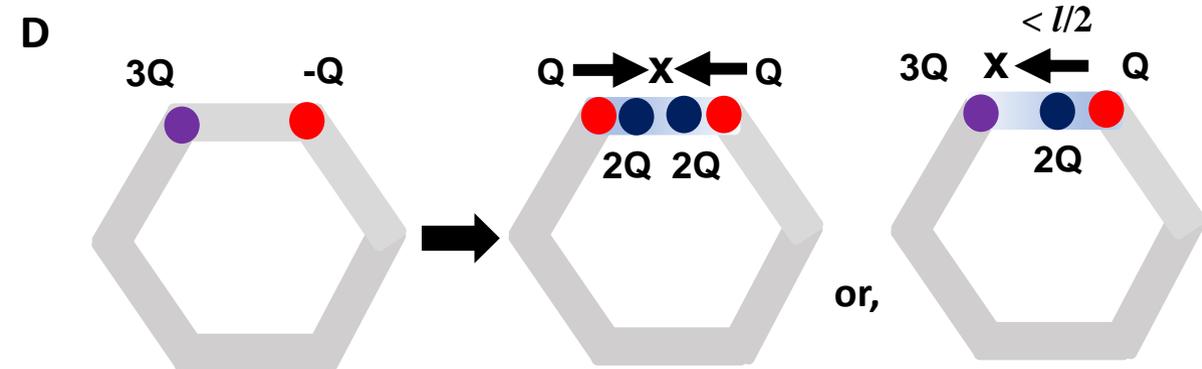
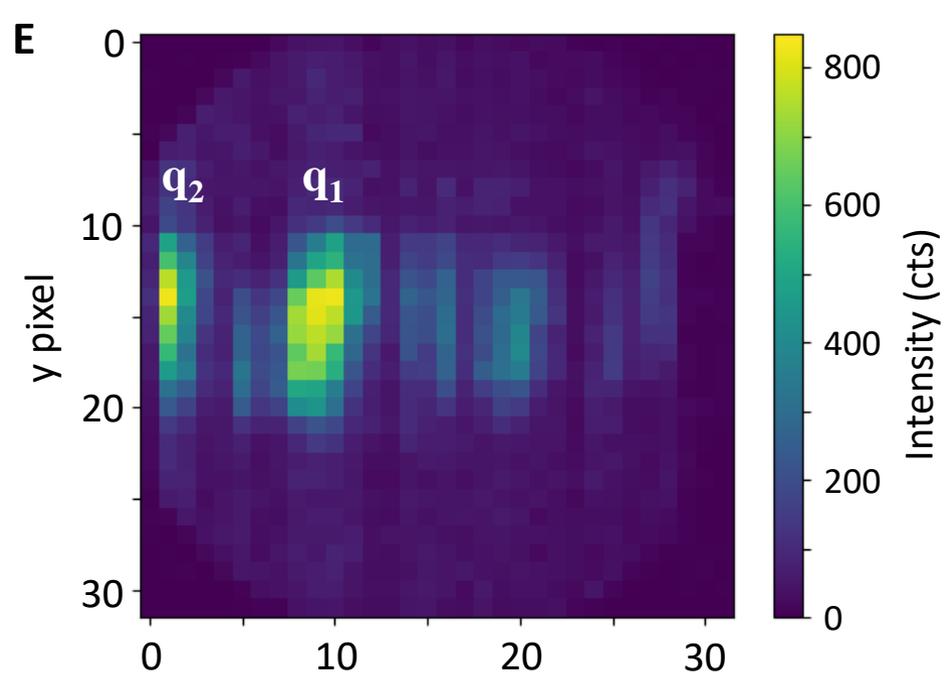
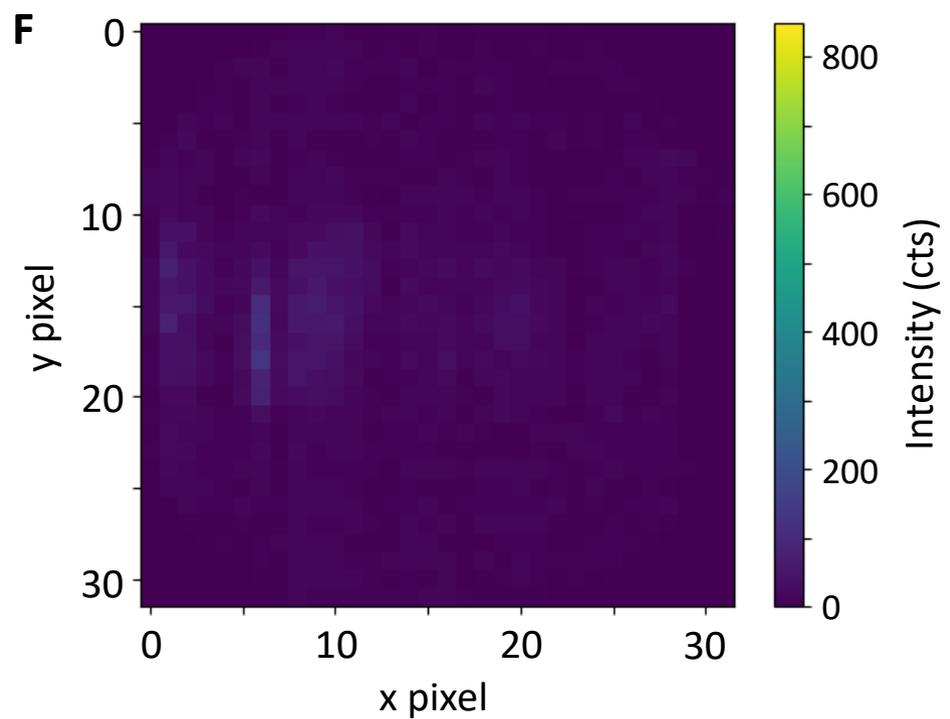

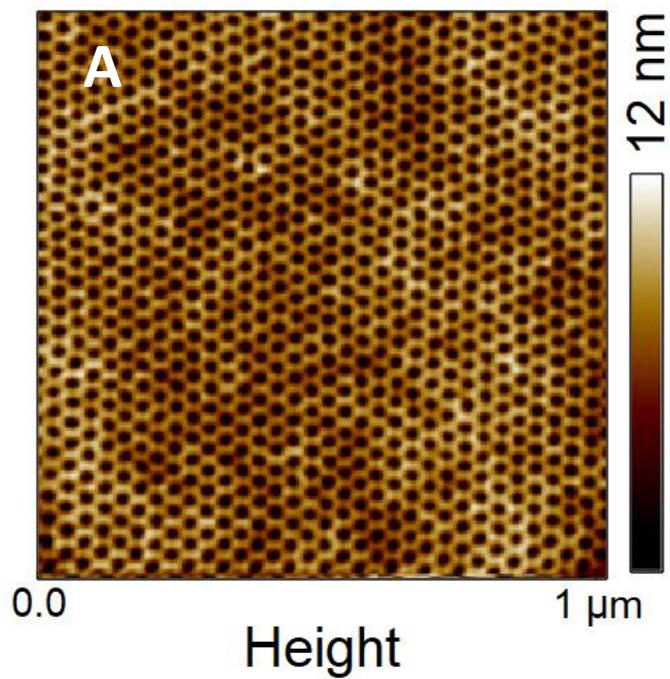
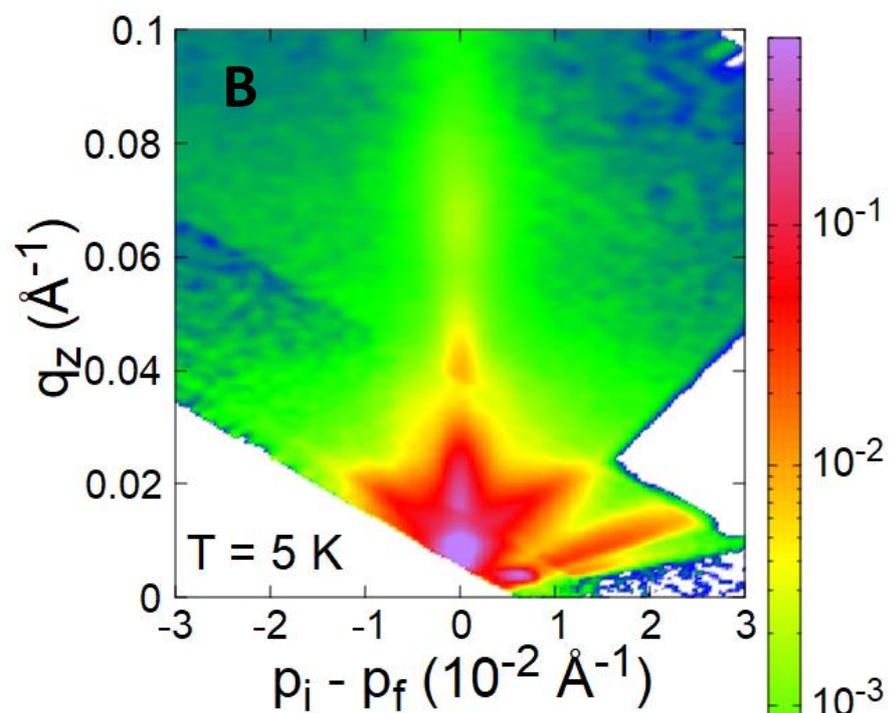
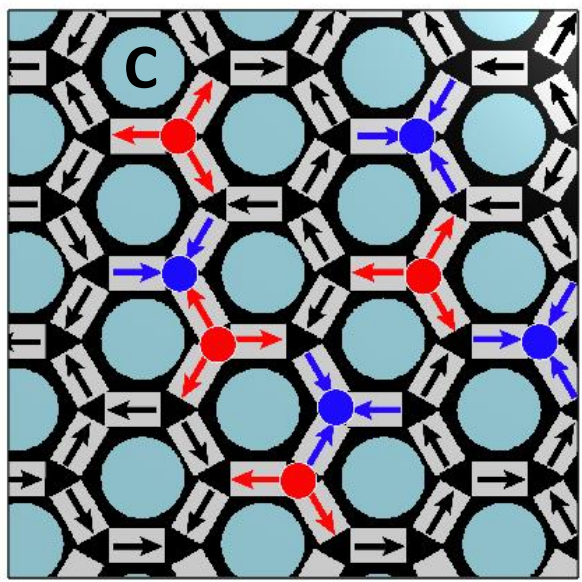
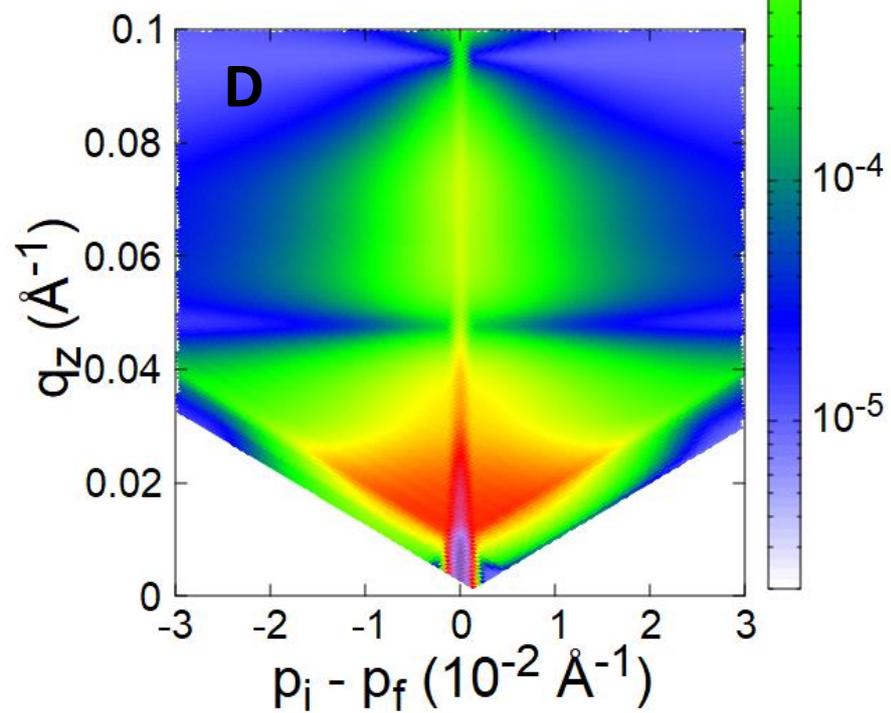

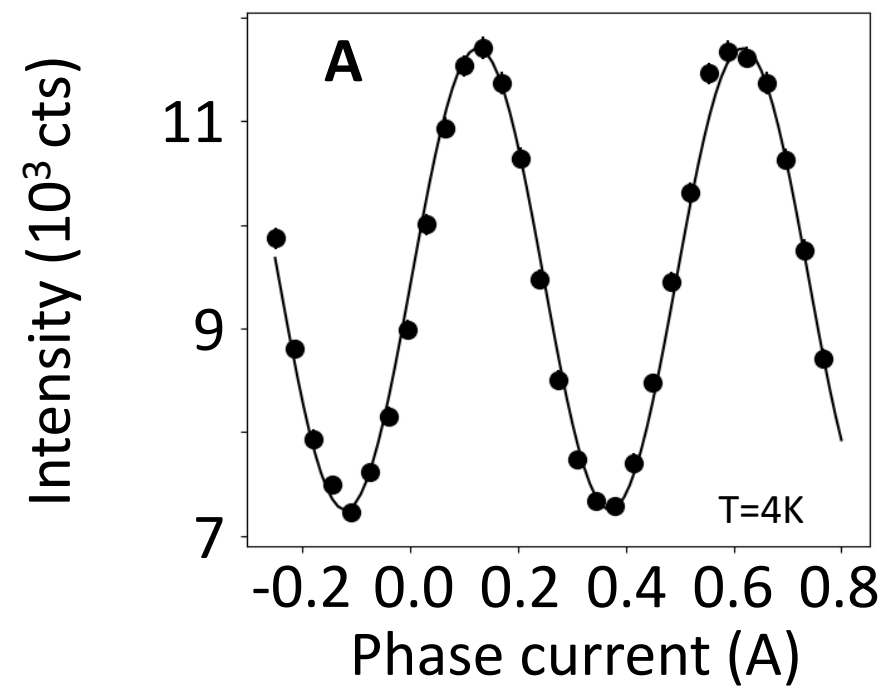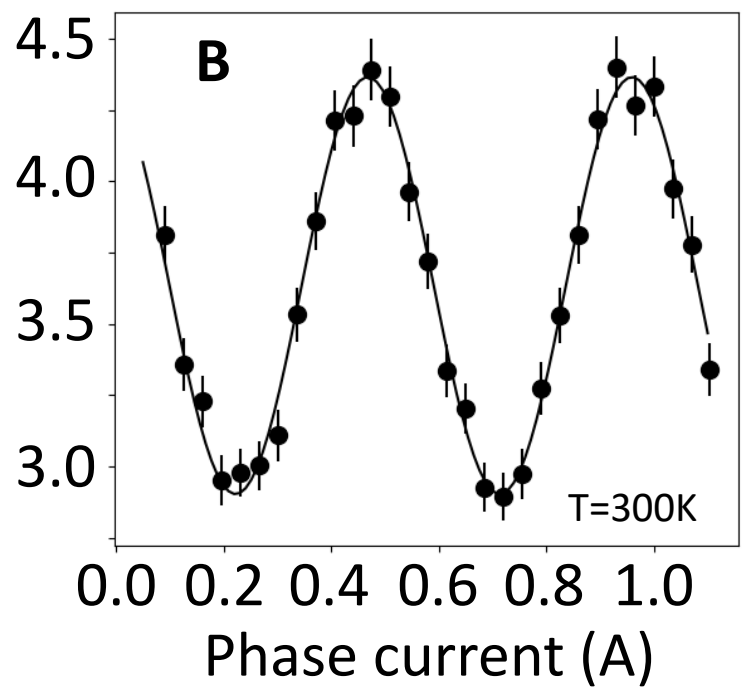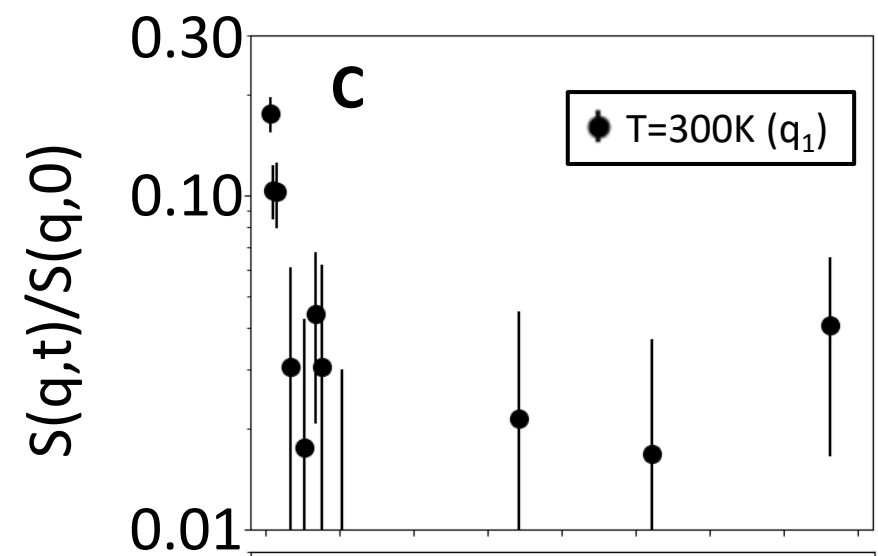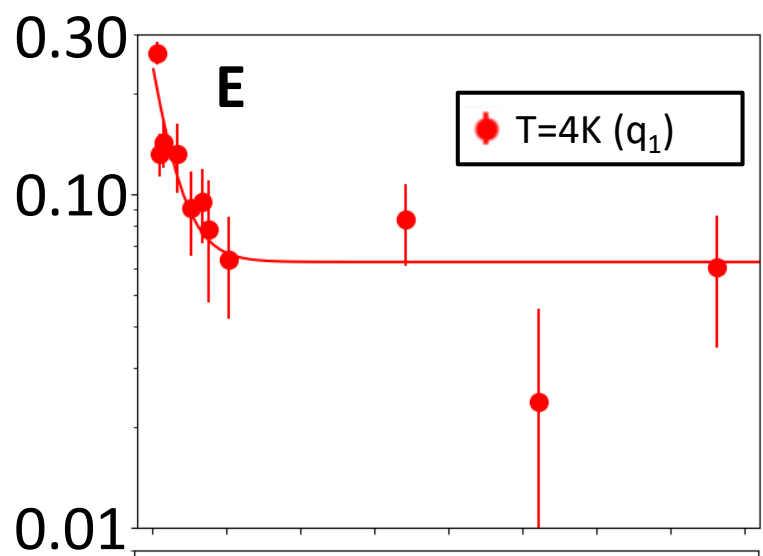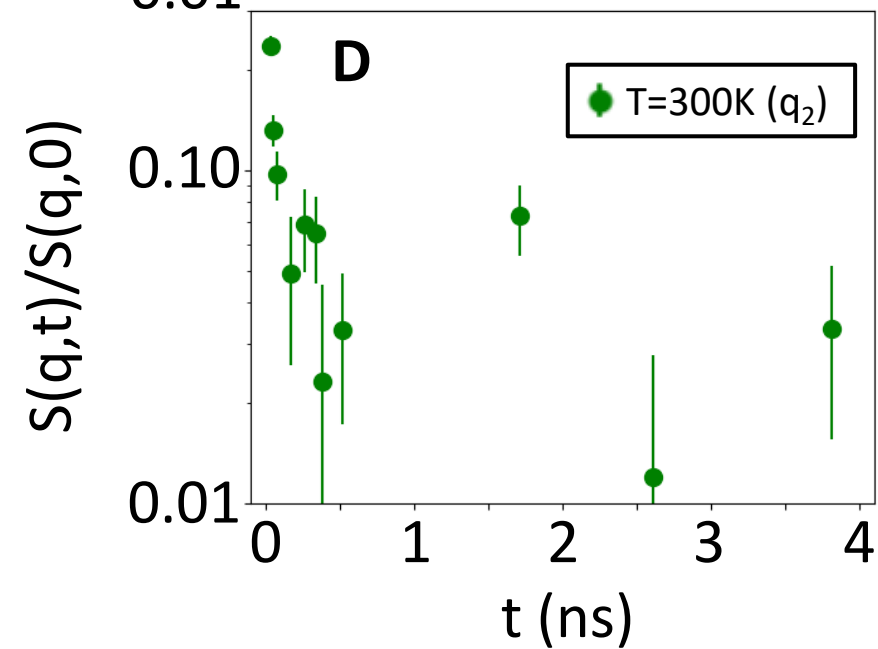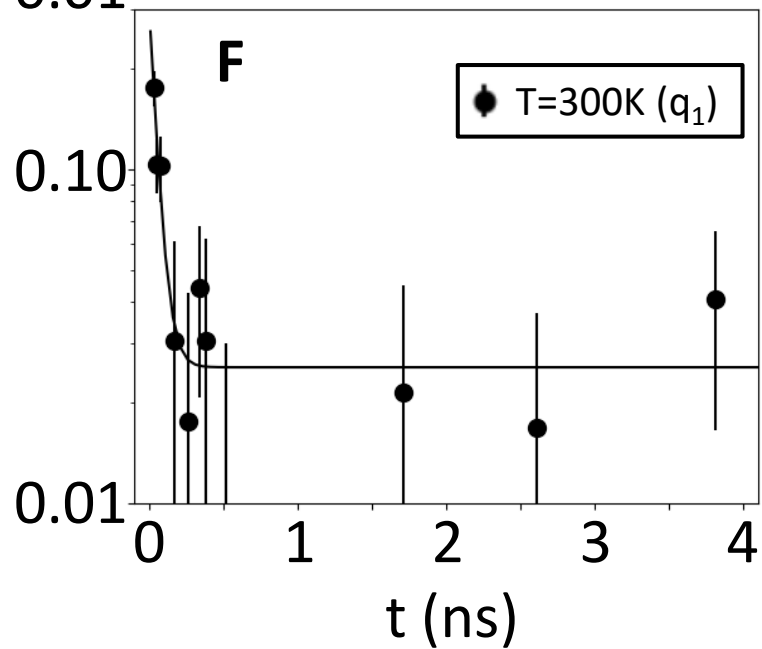

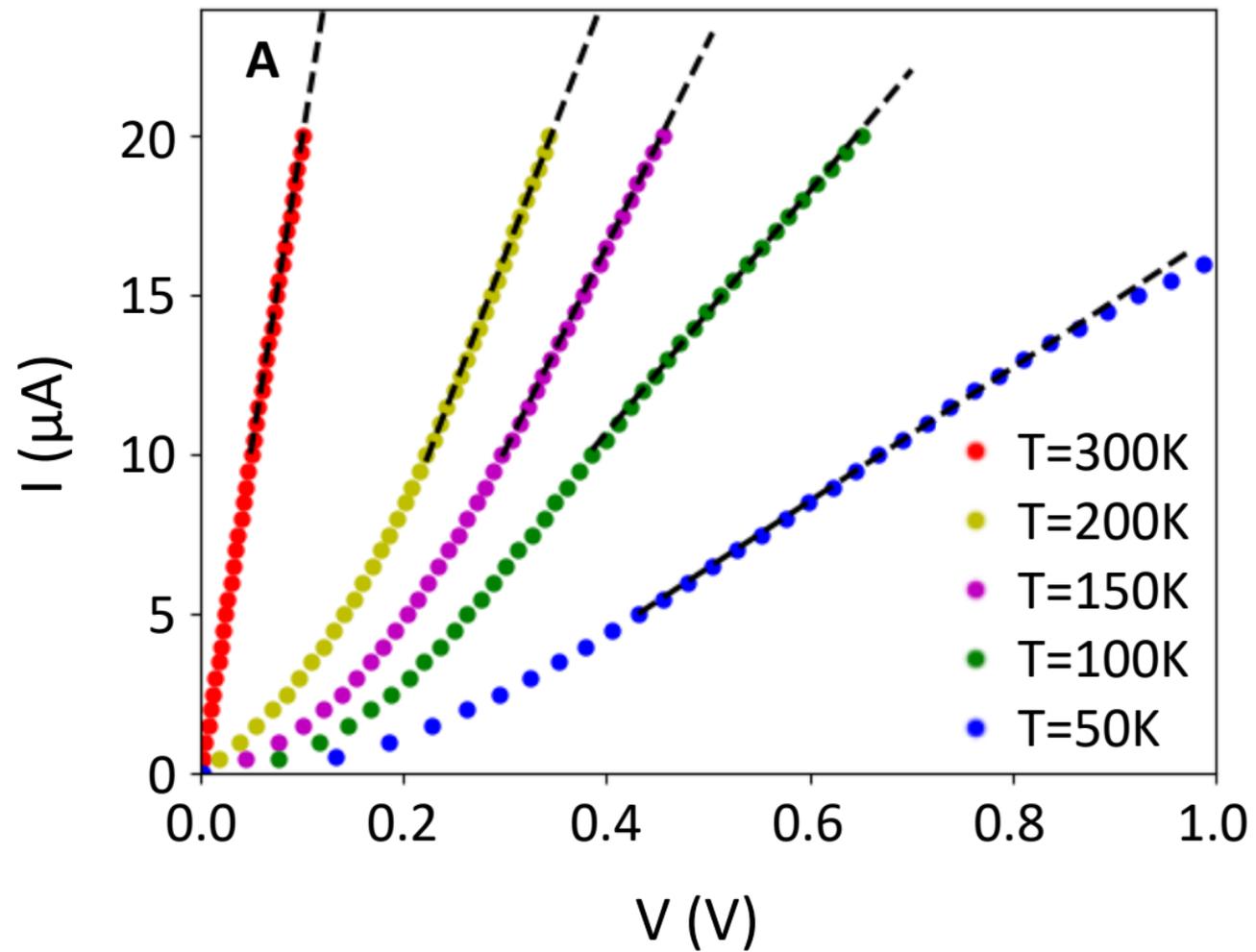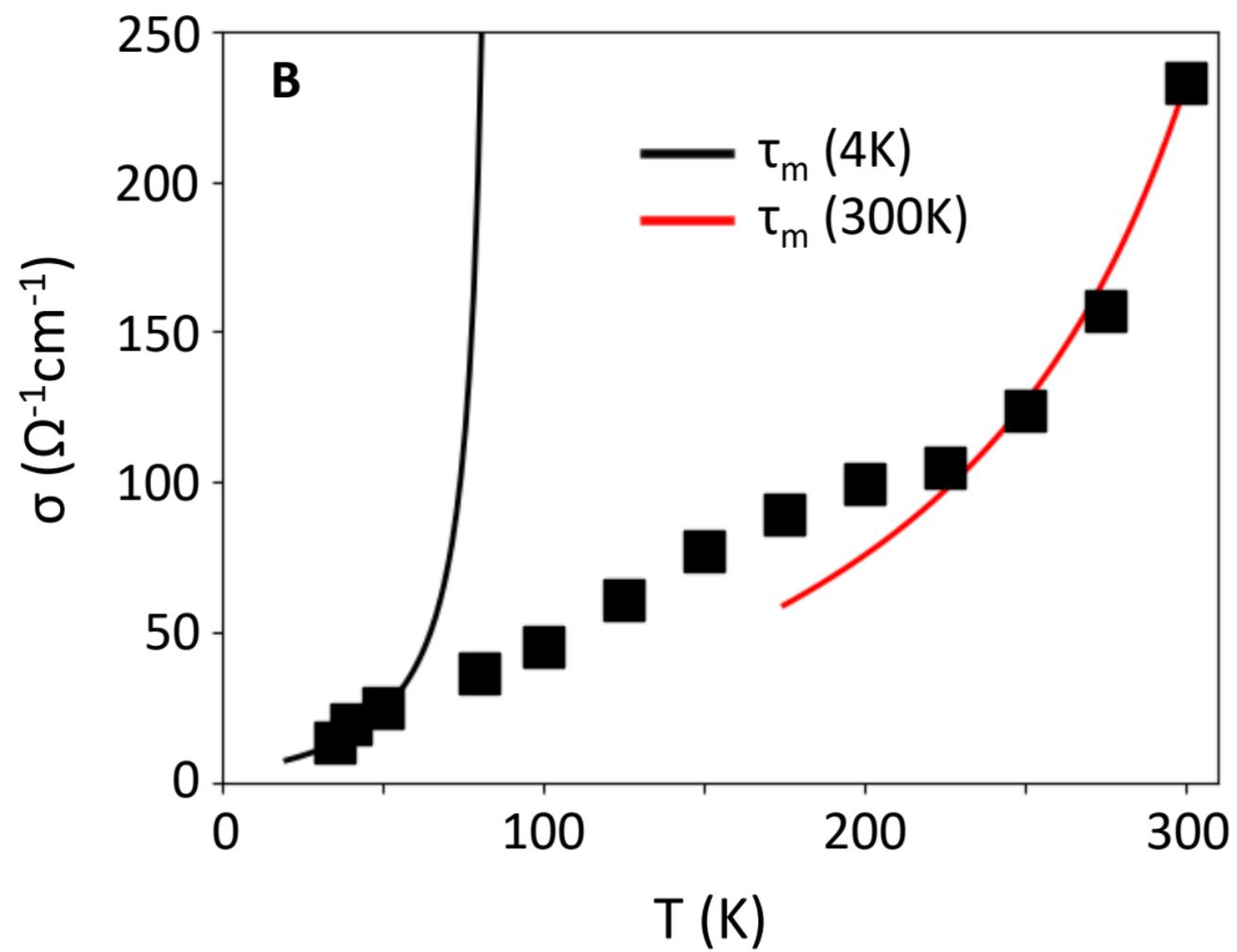